\documentclass[condensedmatter,article,submit,moreauthors]{Definitions/mdpi} 

\firstpage{1} 
\pubvolume{1}
\issuenum{1}
\articlenumber{0}
\pubyear{2026}
\copyrightyear{2026}
\datereceived{ } 
\daterevised{ } % Comment out if no revised date
\dateaccepted{ } 
\datepublished{ } 
\usepackage{tikz}
\usetikzlibrary{decorations.markings}
\usetikzlibrary{decorations.pathmorphing}

\Title{Direct interaction corrections to the density of states in anisotropic conductors with spin–orbit coupling}

\Author{Bahruz Suleymanli $^{1}$\orcidA{} and B. Tanatar $^{2,}$*\orcidB{}}

\AuthorNames{Bahruz Suleymanli and B. Tanatar}

\address{%
$^{1}$ \quad Department of Physics, Yıldız Technical University, 34220 Esenler, Istanbul, Türkiye; bahruz.suleymanli@gmail.com\\
$^{2}$ \quad Department of Physics, Bilkent University, 06800 Ankara, Türkiye; tanatar@fen.bilkent.edu.tr}

\corres{Correspondence: tanatar@fen.bilkent.edu.tr}

\abstract{In this study, we investigate how the direct contribution of electron–electron interactions modifies the density of states anomaly near the Fermi level in a planar array of weakly coupled disordered chains with longitudinal Rashba–Dresselhaus spin–orbit coupling (SOC). The correction shows logarithmic behavior in the two-dimensional regime and an inverse-square-root dependence in the quasi-one-dimensional regime. In both regimes, the direct contribution is positive and partially compensates the negative exchange correction. 
The weight of the direct term relative to the exchange term
depends only weakly on SOC. It is 4\% to 6\% for strong screening and reaches about 20\% for weak screening.
These results suggest that tunneling spectroscopy of weakly screened chain arrays provides favorable conditions for observing the direct contribution.}

\keyword{spin–orbit coupling; density of states; anisotropic conductors} 

\begin{document}

%%%%%%%%%%%%%%%%%%%%%%%%%%%%%%%%%%%%%%%%%%
\section{Introduction}

In disordered conductors, repeated impurity scattering makes electron motion diffusive and increases the time electrons spend interacting with one another \cite{Altshuler1983, Lee1985, Kamenev1999, Bhatt2025}. This interplay between disorder and interactions gives rise to the Altshuler–Aronov (AA) anomaly, which appears as a singular correction to the single-particle density of states (DOS) near the Fermi level \cite{Altshuler1979, Altshuler1980}. The energy dependence of this anomaly strongly reflects the system's effective dimensionality. The Coulomb exchange contribution exhibits a double-logarithmic singularity in two dimensions, whereas in one dimension it follows an inverse-square-root dependence \cite{Altshuler1985, Mora2007, Sas2009}. Experimentally, this behavior can be observed in tunneling measurements as a corresponding zero-bias conductance anomaly \cite{Ossi2013, Liu2014, Carbillet2020, Kuzmiak2023, Im2023}.

The leading interaction correction to the DOS is of first order in the electron–electron interaction and consists of two distinct contributions. The first is the exchange contribution, which involves the small momentum and frequency carried by a diffuson and therefore requires dynamic screening. The second is the direct contribution, which involves the momentum difference between two points on the Fermi surface and can be described by a statically screened interaction averaged over these points. In the screened singlet–triplet description, the spin-dependent interaction correction depends on which triplet diffusion modes remain soft \cite{Finke1983, Zala2001, Gnida2024}. Therefore, when spin dynamics are affected by spin–orbit coupling (SOC), the direct contribution becomes sensitive to which spin channels remain diffusive and must be treated accordingly.

For both the exchange and direct contributions, weakly coupled chains provide one of the clearest settings in which to study the dimensional crossover of the AA anomaly. For this system, the physical origin of the crossover can be understood in terms of the time available for transverse hopping \cite{Firsov1987, Firsov1988, Pfnur2024}. At long times, electrons have enough time to hop between neighboring chains and explore both longitudinal and transverse directions, so their motion becomes effectively two-dimensional. At shorter times, transverse hopping is less effective, and electrons remain mainly confined to a single chain, leading to quasi-one-dimensional behavior. 
Such a crossover picture is also relevant for experimentally realized strongly anisotropic electronic systems. Lateral superlattices, atomic-wire arrays, and, more recently, quasi-one-dimensional electronic systems provide experimental platforms that such anisotropic models can describe \cite{Endo2005,Blumenstein2011,Tegenkamp2012,Park2013,Wang2022,Wu2022}.

Beyond dimensionality, SOC provides an additional mechanism that can modify the interaction-induced DOS correction. It is particularly worth emphasizing that in low-dimensional conductors, SOC can arise from different sources of inversion asymmetry, with the Dresselhaus contribution inherited from bulk inversion asymmetry and the Rashba contribution arising from structural asymmetry associated with confinement or interfaces \cite{Dresselhaus1955, Manchon2015}. These different microscopic origins also make SOC experimentally controllable in several low-dimensional platforms. Gate-tunable SOC has been demonstrated in semiconductor quantum wells and oxide interfaces \cite{Nitta1997, Miller2003, Caviglia2010, Chen2025}, while strong anisotropic SOC has been observed in atomic wires \cite{Tegenkamp2012, Park2013} and tellurium-based systems \cite{Niu2020, Geldiyev2023, Fartab2023}. From the perspective of spin dynamics, an important simplification occurs when the SOC field has a fixed spin axis. In this case, electrons can be classified by their spin projection along this axis, and spin-independent impurity scattering changes their momentum without changing this spin projection \cite{Schliemann2003}. Conservation of this spin projection, however, does not imply that all other spin modes are suppressed, since special SOC symmetries can support long-lived spin-density modes at finite momentum \cite{Bernevig2006}.

In this work, within a model of weakly coupled disordered chains with longitudinal Rashba–Dresselhaus SOC \cite{Suleymanli2026}, we calculate the direct electron–electron interaction contribution to the DOS correction and examine how it complements the exchange contribution across the dimensional crossover. We show that the fixed spin–orbit axis leaves a single triplet channel active, which reduces the direct contribution by a factor of three relative to the spin-degenerate case, while the different branch velocities enhance
it. The paper is organized as follows. Sec.~\ref{sec2} introduces the model, Sec.~\ref{sec:results} presents the results, and
Sec.~\ref{sec:conclusions} summarizes our conclusions. The analytical expressions for the Fermi-surface factors and the diffusion kernel are given in Appendix~\ref{Ap:a}. We use units with $\hbar = k_B =1$ throughout. 

%%%%%%%%%%%%%%%%%%%%%%%%%%%%%%%%%%%%%%%%%%
\section{Spin--orbit coupled diffusive model}
\label{sec2}

We consider a planar array of parallel disordered chains with spacing $d$ and interchain hopping $t_y$, subject to the longitudinal Rashba--Dresselhaus coupling $\alpha k_x\sigma_y-\beta k_x\sigma_x=\lambda k_x\,\hat{\mathbf{n}}\cdot\boldsymbol{\sigma}$, where $\lambda=\sqrt{\alpha^2+\beta^2}$ and $\hat{\mathbf{n}}=(-\beta,\alpha,0)/\lambda$. Since $\hat{\mathbf{n}}$ does not depend on momentum, the spin projection $\sigma=\pm$ on $\hat{\mathbf{n}}$ is conserved and the linearized spectrum has two branches \cite{Suleymanli2026},
\begin{equation}
\varepsilon_\sigma(\mathbf{k})=\varepsilon_F+v_F(|k_x|-k_F)+\sigma\lambda k_x-2t_y\cos(dk_y).
\label{eq:spectrum}
\end{equation}
For $v_F>\lambda$ and $2t_y\ll v_Fk_F$, branch $\sigma$ has velocity $v_F^\sigma=v_F+\sigma\lambda$ on the right Fermi sheet and $v_F^{-\sigma}$ on the left Fermi sheet. Introducing $\gamma=\lambda/v_F$, the SOC increases the separation between the right and left Fermi points to $Q=2k_F/(1-\gamma^2)$. The total DOS is $\rho_0=\sum_\sigma\rho_0^\sigma$, where $\rho_0^\sigma=2/(\pi v_F^\sigma d)$, and the corresponding screening wave vector is $\varkappa=2\pi e^2\rho_0$.
The disorder is spin-independent, with forward and backward scattering times $\tau_1$ and $\tau_2$ and $\tau^{-1}=\tau_1^{-1}+\tau_2^{-1}$, so it changes the momentum of an electron but not its spin projection. For $k_Fl\gg1$, with $l=v_F\tau$, the diffuson of branch $\sigma$ is $[-i\omega+V_1^\sigma(\mathbf{q})]^{-1}$, where longitudinal diffusion is set by backscattering and transverse diffusion by interchain hopping \cite{Firsov1987,Suleymanli2026},
\begin{equation}
V_1^\sigma(\mathbf{q})=D_\parallel^\sigma q_x^2+D_\perp\big[1-\cos(dq_y)\big],
\label{eq:V1}
\end{equation}
where $D_\parallel^\sigma=(v_F^\sigma)^2\tau_2/2,
D_\perp=4t_y^2\tau.$
The crossover energy $\varepsilon_c=2D_\perp=8t_y^2\tau$ separates two-dimensional diffusion at $|\omega|\ll\varepsilon_c$ from quasi-one-dimensional diffusion at $\varepsilon_c\ll|\omega|\ll1/\tau$, and we take $\varepsilon_c\tau\ll1$.

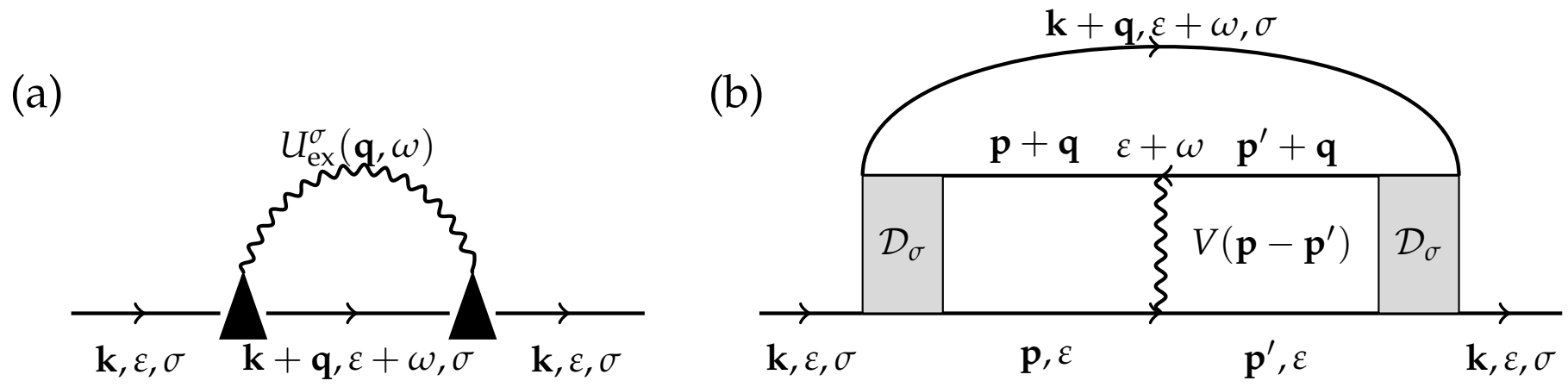
\begin{figure}[tbp]
\centering
\begin{tikzpicture}[scale=0.88,%>=Stealth,
  fermion/.style={thick,postaction={decorate},decoration={markings,mark=at position #1 with {\arrow{>}}}},
  photon/.style={decorate,decoration={snake,amplitude=1.1pt,segment length=4.5pt},thick},
  ladder/.style={fill=gray!30,draw=black,thin}]
% ---------------- (a) exchange
\begin{scope}[xshift=0cm]
\node at (-0.3,1.9) {(a)};
\draw[fermion=0.5] (0,0) -- (1.3,0);
\draw[fermion=0.5] (1.7,0) -- (3.3,0);
\draw[fermion=0.5] (3.7,0) -- (5,0);
\filldraw[black] (1.3,-0.22) -- (1.7,-0.22) -- (1.5,0.35) -- cycle;
\filldraw[black] (3.3,-0.22) -- (3.7,-0.22) -- (3.5,0.35) -- cycle;
\draw[photon] (1.5,0.35) .. controls (1.9,1.5) and (3.1,1.5) .. (3.5,0.35);
\node at (0.6,-0.42) {\footnotesize $\mathbf{k},\varepsilon,\sigma$};
\node at (2.5,-0.42) {\footnotesize $\mathbf{k}+\mathbf{q},\varepsilon+\omega,\sigma$};
\node at (4.4,-0.42) {\footnotesize $\mathbf{k},\varepsilon,\sigma$};
\node at (2.5,1.45) {\footnotesize $U_\mathrm{ex}^\sigma(\mathbf{q},\omega)$};
\end{scope}
% ---------------- (b) direct
\begin{scope}[xshift=6.0cm]
\node at (-0.2,1.9) {(b)};
\draw[fermion=0.5] (0,0) -- (0.9,0);
\draw[thick] (0.9,0) -- (1.6,0);
\draw[fermion=0.5] (1.6,0) -- (5.4,0);
\draw[thick] (5.4,0) -- (6.1,0);
\draw[fermion=0.5] (6.1,0) -- (7.0,0);
\draw[thick] (0.9,1.2) -- (1.6,1.2);
\draw[fermion=0.5] (5.4,1.2) -- (1.6,1.2);
\draw[thick] (5.4,1.2) -- (6.1,1.2);
\draw[fermion=0.5] (0.9,1.2) .. controls (0.9,2.7) and (6.1,2.7) .. (6.1,1.2);
\draw[ladder] (0.9,0.0) rectangle (1.6,1.2);
\draw[ladder] (5.4,0.0) rectangle (6.1,1.2);
\node at (1.25,0.6) {\footnotesize $\mathcal{D}_\sigma$};
\node at (5.75,0.6) {\footnotesize $\mathcal{D}_\sigma$};
\draw[photon] (3.5,0) -- (3.5,1.2);
\node[anchor=west] at (3.62,0.6) {\footnotesize $V(\mathbf{p}-\mathbf{p}')$};
\node at (0.45,-0.4) {\footnotesize $\mathbf{k},\varepsilon,\sigma$};
\node at (2.5,-0.4) {\footnotesize $\mathbf{p},\varepsilon$};
\node at (4.5,-0.4) {\footnotesize $\mathbf{p}',\varepsilon$};
\node at (6.55,-0.4) {\footnotesize $\mathbf{k},\varepsilon,\sigma$};
\node at (2.4,1.45) {\footnotesize $\mathbf{p}+\mathbf{q}$};
\node at (3.5,1.45) {\footnotesize $\varepsilon+\omega$};
\node at (4.6,1.45) {\footnotesize $\mathbf{p}'+\mathbf{q}$};
\node at (3.5,2.5) {\footnotesize $\mathbf{k}+\mathbf{q},\varepsilon+\omega,\sigma$};
\end{scope}
\end{tikzpicture}
\caption{Exchange (\textbf{a}) and direct (\textbf{b}) self-energy diagrams. Triangles and shaded blocks denote impurity ladders. In (\textbf{a}) the interaction line carries the diffuson momentum $\mathbf{q}$ and frequency $\omega$ and is dynamically screened. In (\textbf{b}) it transfers the large momentum $\mathbf{p}-\mathbf{p}'$ between two points of the Fermi surface at zero frequency and is statically screened.}
\label{fig:diagrams}
\end{figure}

To first order in the screened interaction, both corrections to the DOS contain two diffusons and one interaction line (Fig.~\ref{fig:diagrams}) and take the common form \cite{Firsov1987,AGD}
\begin{equation}
\delta\rho_X(E,T)=-\frac{1}{\pi}\sum_\sigma\rho_0^\sigma\int_0^{1/\tau}\!\mathrm{d}\omega\,\Phi_T(E,\omega)\,
\mathrm{Im}\!\int\!\frac{\mathrm{d}^2q}{(2\pi)^2}\,\frac{U_X^\sigma(\mathbf{q},\omega)}{[-i\omega+V_1^\sigma(\mathbf{q})]^2},
\label{eq:master}
\end{equation}
where $X=\mathrm{ex},\mathrm{dir}$ labels the channel and $\Phi_T(E,\omega)=\tfrac12\big[\tanh\tfrac{\omega+E}{2T}+\tanh\tfrac{\omega-E}{2T}\big]$ reduces to $\theta(\omega-|E|)$ at $T=0$. In the exchange channel the interaction carries the small diffuson momentum and frequency and is dynamically screened, $U_\mathrm{ex}^\sigma=2\pi e^2(-i\omega+V_1^\sigma)/(-i\omega|\mathbf{q}|+\varkappa^\sigma V_1^\sigma)$ with $\varkappa^\sigma=2\pi e^2\rho_0^\sigma$ \cite{Suleymanli2026}. In the direct channel, it transfers the large momentum $\mathbf{p}-\mathbf{p}'$ between two Fermi-surface points at zero frequency, so it enters through the static interaction $V(\mathbf{q})=2\pi e^2/(|\mathbf{q}|+\varkappa)$ \cite{Firsov1987}. The additional fermion loop makes the direct correction positive.

The fixed SOC axis also determines the spin channels contributing to the direct term. Each closed loop preserves a single spin projection, while ladders connecting opposite spin projections are suppressed by the spin-orbit energy splitting $\Delta_\mathrm{SO}=2\lambda k_F$. Consequently, for $\Delta_\mathrm{SO}\tau\gg1$, only same-spin ladders retain a diffusion pole at small $\mathbf{q}$. Long-lived mixed-spin modes at finite momentum, which may occur for special SOC symmetries \cite{Bernevig2006}, are not considered. In the singlet--triplet language \cite{Altshuler1985,Zala2001}, the density part of the same-spin amplitude is absorbed into the screened exchange term for $q\ll\varkappa$, and the helicity-density part gives the constant amplitude
\begin{equation}
U_\mathrm{dir}=-\frac{1}{2}\big\langle V(\mathbf{p}-\mathbf{p}')\big\rangle\equiv-\frac{F}{\rho_0}.
\label{eq:Udir}
\end{equation}
Backscattering places the far end of each ladder on either sheet with equal probability, so the Fermi-surface average splits into intrasheet and intersheet parts of equal weight, $F=F_a+F_b$, with 
\begin{equation}
F_j=\frac{\varkappa d^2}{4}\iint_\mathrm{BZ}\frac{\mathrm{d}k\,\mathrm{d}k'}{(2\pi)^2}\,\frac{1}{\sqrt{Q_j^2+(k-k')^2}+\varkappa},\qquad Q_a=0,\quad Q_b=Q,
\label{eq:FaFb}
\end{equation}
where $k,k'\in(-\pi/d,\pi/d)$ are transverse momenta.

Since $U_\mathrm{dir}$ is constant, Eq.~(\ref{eq:master}) reduces for the direct channel to
\begin{equation}
\begin{gathered}
\delta\rho_\mathrm{dir}(E,T)=\frac{F}{\pi\rho_0}\sum_\sigma\rho_0^\sigma\int_0^{1/\tau}\!\mathrm{d}\omega\,\Phi_T(E,\omega)\,K_\sigma(\omega),\\
K_\sigma(\omega)=\mathrm{Im}\!\int\!\frac{\mathrm{d}^2q}{(2\pi)^2}\,\frac{1}{[-i\omega+V_1^\sigma(\mathbf{q})]^2}.
\end{gathered}
\label{eq:dDir}
\end{equation}
The interaction thus enters only through $F$, and the entire energy dependence is carried by the two-diffuson kernel $K_\sigma(\omega)$, which we evaluate in Sec.~\ref{sec:results}.

\section{Results}
\label{sec:results}

The SOC modifies the direct contribution by selecting spin channels, changing the Fermi-surface factor $F$ via screening and intersheet momentum transfer, and altering the diffusion kernel via different branch velocities. Below, we examine these effects and compare the direct contribution with the exchange correction of Ref.~\cite{Suleymanli2026}.

\subsection{Spin channels selected by the spin--orbit coupling}
\label{sec:spin}

In a spin-degenerate array, the density channel is included in the screened exchange term, while the three triplet components contribute equally, giving the total amplitude $-\tfrac32\langle V\rangle$ \cite{Altshuler1985,Zala2001}. The fixed SOC axis changes this result. Only the triplet component with zero spin projection along $\hat{\mathbf n}$ survives as the helicity-density channel of Eq.~(\ref{eq:Udir}). The other two components involve opposite spin projections and are detuned by $\Delta_\mathrm{SO}$. Consequently, the direct contribution is reduced by a factor of three compared with the spin-degenerate case.

This reduction does not involve spin relaxation and is analogous to the suppression of two triplet diffusons by a Zeeman field \cite{Altshuler1985}. In contrast, spin--orbit scattering suppresses all three triplet components. 
A weak Zeeman field, $g\mu_BB\ll\Delta_\mathrm{SO}$, therefore has little effect on the direct contribution. The reduction holds for $\Delta_\mathrm{SO}\tau=2\gamma k_Fl\gg1$, while at weaker SOC the opposite-spin channels gradually recover their contribution. Long-lived mixed-spin modes at finite momentum \cite{Bernevig2006} are not considered here.

\subsection{Fermi-surface factors}
\label{sec:factors}

\begin{figure}[tbp]
\centering
\includegraphics[width=\textwidth]{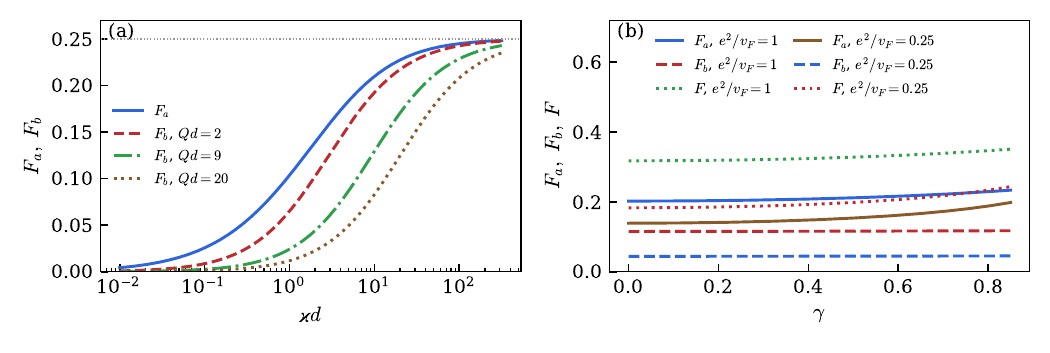}
\caption{(\textbf{a}) Intrasheet and intersheet factors as functions of $\varkappa d$ for different $Qd$. (\textbf{b}) SOC dependence for $k_Fd=4.5$.}
\label{fig:F}
\end{figure}

The factor $F$ describes the statically screened interaction at the relevant Fermi-surface momentum transfers. Fig.~\ref{fig:F}(a) shows its intrasheet and intersheet components as functions of $\varkappa d$. Both increase with screening and approach $1/4$ in the strong-screening limit. The intersheet factor $F_b$ is smaller because of the large momentum transfer $Q\simeq2k_F$. For $\varkappa\ll Q$ and $Qd\gg2\pi$, it reduces to $F_b\simeq\varkappa/4Q$. In the weak-screening limit, $\varkappa d\ll1$, the intrasheet factor behaves as $F_a\simeq(\varkappa d/4\pi)\ln(2\pi/\varkappa d)$. The closed expressions for both factors are given in Appendix~\ref{Ap:a}.

The SOC enters these factors through $\varkappa$ and $Q$, both proportional to $(1-\gamma^2)^{-1}$. Their ratio is therefore independent of $\gamma$, and $F$ is the same for both branches. Fig.~\ref{fig:F}(b) shows that the SOC increases $F$, although the effect is relatively small. For $e^2=v_F$ and $k_Fd=4.5$, we obtain $F_a=0.202$, $F_b=0.115$, and $F=0.317$ at $\gamma=0$. The total factor increases to $F=0.331$ and $0.339$ at $\gamma=0.57$ and $0.70$, respectively. For $e^2=v_F/4$, the corresponding value at $\gamma=0$ is $F=0.18$. At $\gamma=0.7$, the enhancement reaches $7\%$ and $19\%$ for the stronger and weaker Coulomb couplings, respectively. Thus, the main SOC dependence of the direct correction comes from the diffusion kernel rather than $F$.

\subsection{Energy dependence across the dimensional crossover}
\label{sec:energy}

The kernel $K_\sigma(\omega)$ in Eq.~(\ref{eq:dDir}) describes the diffusive return probability and determines the energy dependence of the direct correction. After integrating over the longitudinal momentum, the transverse integral can be expressed through a complete elliptic integral, as shown in Appendix~\ref{Ap:a}. The remaining frequency dependence is determined by $\omega/\varepsilon_c$, and the two branch contributions can be summed exactly. At $T=0$, we obtain

\begin{equation}
\delta\rho_\mathrm{dir}(E)=A_\mathrm{dir}\Big[\mathcal{H}\Big(\frac{|E|}{\varepsilon_c}\Big)-\mathcal{H}\Big(\frac{1}{\varepsilon_c\tau}\Big)\Big],
\label{eq:dDirH}
\end{equation}
where $A_\mathrm{dir}=F\,(1+\gamma^2)/\left(8\pi v_Fd\,t_y\sqrt{\tau\tau_2}\,(1-\gamma^2)\right)$. The second term accounts for the upper cutoff $1/\tau$ and becomes small for $\varepsilon_c\tau\ll1$, since $\mathcal{H}(1/\varepsilon_c\tau)\simeq\sqrt{2\varepsilon_c\tau}$. At finite temperature and $|E|\ll T$, the energy scale $|E|$ is replaced by $T$ up to a numerical factor of order unity.

\begin{figure}[tbp]
\centering
\includegraphics[width=1\textwidth]{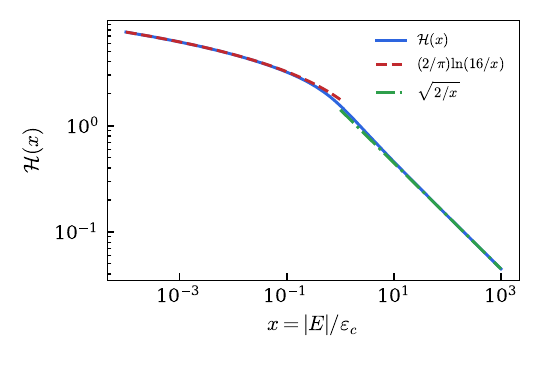}
\caption{Crossover function $\mathcal{H}(x)$ and its two-dimensional and quasi-one-dimensional asymptotes.}
\label{fig:H}
\end{figure}

The function $\mathcal{H}(x)$ is universal, with the SOC, disorder, and interaction entering only through $A_\mathrm{dir}$ and $\varepsilon_c$. As shown in Fig.~\ref{fig:H}, it approaches $(2/\pi)\ln(16/x)$ for $x\ll1$ and $\sqrt{2/x}$ for $x\gg1$. These limits give

\begin{equation}
\delta\rho_\mathrm{dir}(E)\simeq
\begin{cases}
\dfrac{F\,(1+\gamma^2)}{4\pi^2v_Fd\,t_y\sqrt{\tau\tau_2}\,(1-\gamma^2)}\,\ln\dfrac{16\varepsilon_c}{|E|}, & |E|\ll\varepsilon_c,\\[14pt]
\dfrac{F\,(1+\gamma^2)}{2\pi v_Fd\sqrt{\tau_2}\,(1-\gamma^2)}\,\dfrac{1}{|E|^{1/2}}, & \varepsilon_c\ll|E|\ll1/\tau.
\end{cases}
\label{eq:limits}
\end{equation}
In the two-dimensional regime, the direct correction grows logarithmically, whereas the exchange correction exhibits a stronger squared-logarithmic singularity due to dynamical screening. In the quasi-one-dimensional regime, the direct correction follows a positive $|E|^{-1/2}$ law, with an amplitude independent of the interchain hopping. At these energies, electrons effectively diffuse along individual chains.

\subsection{Spin--orbit enhancement of the direct correction}
\label{sec:soc}

The prefactor $A_\mathrm{dir}$ contains the SOC-dependent factor $(1+\gamma^2)/(1-\gamma^2)$, which originates from the different branch velocities. The DOS of each branch is proportional to $1/v_F^\sigma$, while its return probability scales as $(D_\parallel^\sigma)^{-1/2}\propto1/v_F^\sigma$. Consequently, the slower branch gives a larger contribution. Summing over both branches yields $\sum_\sigma(v_F^\sigma)^{-2}=2(1+\gamma^2)/[v_F^2(1-\gamma^2)^2]$. After normalization by the total DOS, $\rho_0\propto(1-\gamma^2)^{-1}$, we obtain the factor $(1+\gamma^2)/(1-\gamma^2)$. Together with the increase of $F$, this leads to an enhancement of the direct correction.

\begin{figure}[tbp]
\centering
\includegraphics[width=\textwidth]{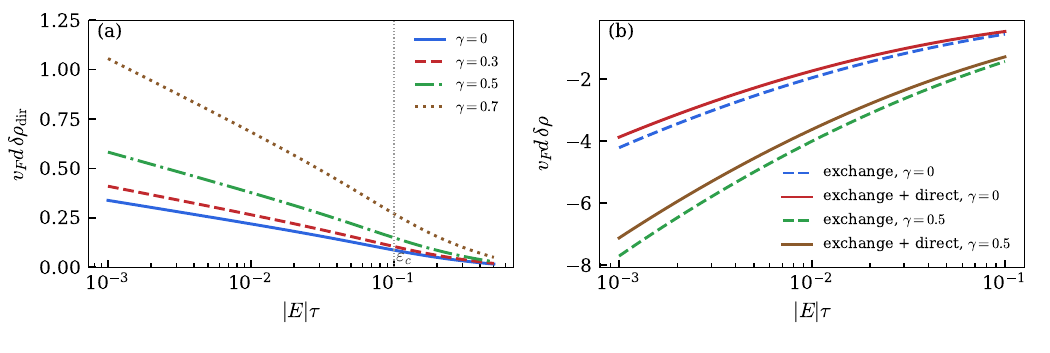}
\caption{(\textbf{a}) Direct DOS correction for different SOC strengths. The dotted line marks $\varepsilon_c$. (\textbf{b}) Exchange correction (dashed) and total correction (solid) for $\gamma=0,\,0.5$. Parameters are $e^2=v_F$, $d=l$, $\tau_2=2\tau$, $\varepsilon_c\tau=0.1$, and $k_Fd=4.5$.}
\label{fig:dos}
\end{figure}

Fig.~\ref{fig:dos}(a) shows the direct correction for the parameters of Ref.~\cite{Suleymanli2026}. It is positive and changes from logarithmic behavior to the $|E|^{-1/2}$ law above $\varepsilon_c$. At $|E|=10^{-3}/\tau$, the dimensionless correction $v_Fd\,\delta\rho_\mathrm{dir}$ increases from $0.34$ at $\gamma=0$ to $1.06$ at $\gamma=0.7$. This nearly threefold enhancement is mainly due to the velocity factor, which reaches $2.92$, while $F$ increases by only $7\%$.

\subsection{Direct versus exchange contribution}
\label{sec:compare}

We now combine the direct correction with the exchange result of Ref.~\cite{Suleymanli2026}. In the two-dimensional regime, the exchange correction reads \cite{Suleymanli2026}
\begin{equation}
\delta\rho_{\mathrm{ex}}^{(\mathrm{2D})}(E)=-\frac{\ln(\tau|E|)}{8\pi^2d\sqrt{\tau\tau_2}\,t_y\,(v_F^2-\lambda^2)}
\Big\{v_F\big[\ln(\tau|E|)+4\Lambda_1\big]+\lambda\Lambda_2\Big\},
\label{eq:ex2D}
\end{equation}
with the constants
\begin{equation}
\begin{gathered}
\Lambda_1=\ln\frac{v_F^2-\lambda^2+4d\sqrt{\tau/\tau_2}\,t_y\big(v_F+d\sqrt{\tau/\tau_2}\,t_y\big)}{128\,e^4\tau^2t_y^2},\\
\Lambda_2=4\ln\frac{v_F-\lambda+d\sqrt{\tau/\tau_2}\,t_y}{v_F+\lambda+d\sqrt{\tau/\tau_2}\,t_y}.
\end{gathered}
\label{eq:Lambda12}
\end{equation} 
The direct contribution has the same prefactor and adds only to the single-logarithmic term. Using $\ln(16\varepsilon_c/|E|)=-\ln(\tau|E|)+\ln(16\varepsilon_c\tau)$, the total correction retains the form of Eq.~(\ref{eq:ex2D}), apart from an energy-independent constant, with the replacement

\begin{equation}
\Lambda_1\;\to\;\Lambda_1+\frac{F\,(1+\gamma^2)}{2}.
\label{eq:Lambda1}
\end{equation}
Thus, the direct contribution does not change the leading $\ln^2$ singularity but reduces the depth of the DOS anomaly. The shift increases with SOC, while $\Lambda_1$ decreases with $\gamma$ \cite{Suleymanli2026}. Fig.~\ref{fig:dos}(b) shows that the relative reduction remains nearly independent of SOC because both corrections contain the common factor $(v_F^2-\lambda^2)^{-1}$.
In the quasi-one-dimensional regime, both corrections follow the $|E|^{-1/2}$ law. The exchange term is negative \cite{Suleymanli2026},
\begin{equation}
\delta\rho_{\mathrm{ex}}^{(\mathrm{1D})}(E)=-\frac{4e^2\Lambda_3}{d\,(v_F^2-\lambda^2)^2\,|\tau E|^{1/2}},
\label{eq:ex1D}
\end{equation}
with
\begin{equation}
\Lambda_3=(v_F^2+\lambda^2)\Big[2+\ln\frac{\pi^2(v_F^2-\lambda^2)}{16e^4}\Big]+2v_F\lambda\ln\frac{v_F-\lambda}{v_F+\lambda},
\label{eq:Lambda3}
\end{equation}
while the direct term is positive. Their sum therefore retains the same energy dependence but has a reduced amplitude.

\begin{figure}[tbp]
\centering
\includegraphics[width=1\textwidth]{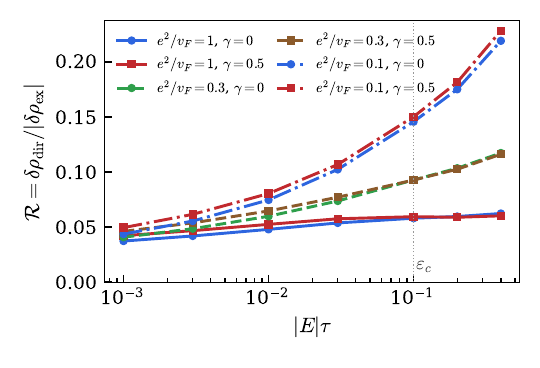}
\caption{Ratio $\mathcal{R}(E)=\delta\rho_\mathrm{dir}/|\delta\rho_\mathrm{ex}|$ for $e^2/v_F=1,\,0.3,\,0.1$ and two SOC strengths. The dotted line marks $\varepsilon_c$.}
\label{fig:ratio}
\end{figure}

Fig.~\ref{fig:ratio} shows the relative contribution $\mathcal{R}(E)=\delta\rho_\mathrm{dir}/|\delta\rho_\mathrm{ex}|$, calculated by numerical integration of both kernels in Eq.~(\ref{eq:master}) without asymptotic approximations. For strong screening, $\varkappa_0d=8$, the direct term amounts to about $4\%$ to $6\%$ of the exchange correction, consistent with Ref.~\cite{Suleymanli2026}. In this regime, $F_a$ approaches its saturation value, while the exchange amplitude continues to increase as $(\varkappa d)^{1/2}$.
For weak screening, $\varkappa_0d\lesssim1$, the exchange amplitude becomes proportional to $\varkappa d\ln(\pi/\varkappa d)$, which has the same screening dependence as $F_a$. Consequently, $\mathcal{R}$ increases to about $15\%$ to $20\%$ in the quasi-one-dimensional regime. The ratio also increases across the dimensional crossover. At low energies, the additional logarithm in the exchange term gives $\mathcal{R}\propto1/\ln(1/|E|\tau)$, whereas above $\varepsilon_c$ both contributions follow the same $|E|^{-1/2}$ law and their ratio approaches a constant.

The SOC dependence of $\mathcal{R}$ is relatively weak because the velocity factors largely cancel between the direct and exchange contributions. This explains the close agreement between the curves for $\gamma=0$ and $\gamma=0.5$ in Fig.~\ref{fig:ratio}. However, the spin selection discussed in Sec.~\ref{sec:spin} reduces the relative weight of the direct term by a factor of three compared with the spin-degenerate case.

These results can be examined using tunneling spectroscopy in SOC-induced chain arrays, including atomic wires on semiconductor surfaces and tellurium-based systems \cite{Blumenstein2011,Tegenkamp2012,Park2013,Niu2020,Geldiyev2023}. The direct contribution reduces the zero-bias DOS dip, particularly in weakly screened systems and at energies above $\varepsilon_c$. In the quasi-one-dimensional regime, it reduces the amplitude of the $|eV|^{-1/2}$ anomaly by a factor $1-\mathcal{R}$, with $\mathcal{R}$ reaching approximately $0.2$.

\subsection{Range of validity}
\label{sec:validity}

Our results are obtained to first order in the screened interaction and leading order in $1/k_Fl$, using Eq.~(\ref{eq:master}). They require $|E|\tau\ll1$ and, for the direct contribution, $\Delta_\mathrm{SO}\tau\gg1$ and $q\ll\varkappa$. The latter condition allows the density channel to be included in the screened exchange term. Following Ref.~\cite{Firsov1987}, the intersheet interaction is treated within static Thomas--Fermi screening. The $2k_F$ structure of the one-dimensional polarization, suppressed by interchain tunneling and disorder, may introduce logarithmic corrections to $F_b$. Close to the Fermi level, the perturbative exchange correction requires resummation \cite{Kamenev1999}. This does not change our conclusions in the two-dimensional regime, where the direct term is subleading. In the quasi-one-dimensional regime, both contributions have the same energy dependence and are treated on an equal footing.

\section{Conclusions}
\label{sec:conclusions}

We have calculated the direct contribution to the Altshuler--Aronov correction of the DOS in weakly coupled disordered chains with Rashba--Dresselhaus SOC, complementing the exchange results of Ref.~\cite{Suleymanli2026}. The SOC suppresses two spin channels, reducing the direct contribution by a factor of three compared with the spin-degenerate case. At the same time, it enhances the correction mainly through the different branch velocities, giving the factor $(1+\gamma^2)/(1-\gamma^2)$, while its effect through the Fermi-surface factor $F$ is small.

The energy dependence is described by a universal crossover function obtained from the exact evaluation of the two-diffuson kernel. In the two-dimensional regime, the direct correction is logarithmic and modifies only the subleading term of the exchange contribution. In the quasi-one-dimensional regime, it follows a positive $|E|^{-1/2}$ law and reduces the amplitude of the exchange singularity. Its relative contribution is weakly dependent on SOC and increases from about $4$--$6\%$ for strong screening to $20\%$ for weak screening. These results suggest that tunneling spectroscopy of weakly screened chain arrays above the crossover energy is a suitable means of observing the direct contribution. 

%%%%%%%%%%%%%%%%%%%%%%%%%%%%%%%%%%%%%%%%%%

%%%%%%%%%%%%%%%%%%%%%%%%%%%%%%%%%%%%%%%%%%
\vspace{6pt} 

\authorcontributions{B.S.: Conceptualization, Formal analysis, Writing -- original draft. B.T.: Conceptualization, Supervision, Writing -- review and editing.}

\funding{This work is supported in part by the Turkish Academy of Sciences (TUBA) under Grant No. AD-2026.}

\institutionalreview{Not applicable.}

\informedconsent{Not applicable.}

\dataavailability{The results supporting the findings of this study are presented in the article. Further information is available from the corresponding author upon reasonable request.} 

\conflictsofinterest{The authors declare no conflicts of interest.} 

%%%%%%%%%%%%%%%%%%%%%%%%%%%%%%%%%%%%%%%%%%
\abbreviations{Abbreviations}{%
The following abbreviations are used in this manuscript:\\

\noindent 
\begin{tabular}{@{}ll}
AA & Altshuler–Aronov\\
DOS & density of states\\
SOC & spin–orbit coupling
\end{tabular}
}

%%%%%%%%%%%%%%%%%%%%%%%%%%%%%%%%%%%%%%%%%%
%% Optional
\appendixtitles{no} % Leave argument "no" if all appendix headings stay EMPTY (then no dot is printed after "Appendix A"). If the appendix sections contain a heading, then change the argument to "yes".
\appendixstart
\appendix
\section[\appendixname~\thesection]{}\label{Ap:a}

The double integrals in Eq.~(\ref{eq:FaFb}) depend only on $u=k-k'$. With $L=2\pi/d$, they reduce to single integrals,
\begin{equation}
F_j=\frac{\varkappa d^2}{8\pi^2}\int_0^L\!\mathrm{d} u\,\frac{L-u}{\sqrt{Q_j^2+u^2}+\varkappa}.
\label{eq:Fsingle}
\end{equation}
The intrasheet factor is 
\begin{equation}
F_a=\frac12\Big(\frac{\varkappa d}{2\pi}\Big)^2\Big[(1+x)\ln(1+x)-x\Big],
\label{eq:Fa}
\end{equation}
where $x=2\pi/\left(\varkappa d\right).$
The intersheet factor can be evaluated analytically for arbitrary $\varkappa/Q$, giving
\begin{equation}
F_b=\frac{\varkappa d^2}{8\pi^2}\Big[L\,\mathrm{arsinh}\frac{L}{Q}+Q-r_L+\varkappa\ln\frac{r_L+\varkappa}{Q+\varkappa}+\varkappa L\,\mathcal T\Big],\qquad r_L=\sqrt{Q^2+L^2},
\label{eq:Fb}
\end{equation}
with
\begin{equation}
\mathcal T=\begin{cases}
\dfrac1A\Big[\arctan\dfrac{\varkappa L}{A\,r_L}-\arctan\dfrac{L}{A}\Big], & A=\sqrt{Q^2-\varkappa^2},\quad \varkappa<Q,\\[10pt]
\dfrac{1}{2B}\ln\left|\dfrac{(\varkappa L-B\,r_L)(L+B)}{(\varkappa L+B\,r_L)(L-B)}\right|, & B=\sqrt{\varkappa^2-Q^2},\quad \varkappa>Q,
\end{cases}
\label{eq:T}
\end{equation}
which is continuous at $\varkappa=Q$. For $\varkappa\ll Q$ this reduces to Eq.~(29) of Ref.~\cite{Firsov1987}, and for $\varkappa\gg Q$ the intersheet factor approaches the intrasheet one. 

The longitudinal integral in the kernel of Eq.~(\ref{eq:dDir}) gives $K_\sigma(\omega)=(D_\parallel^\sigma)^{-1/2}\,\mathrm{Im}\ J(\omega)/\left(4d\right)$. The remaining transverse integral is a complete elliptic integral of the second kind,
\begin{equation}
J(\omega)=\frac{1}{2\pi}\int_{-\pi}^{\pi}\!\mathrm{d}\varphi\,\big[-i\omega+D_\perp(1-\cos\varphi)\big]^{-3/2}
=\frac{2}{\pi}\,\frac{(\varepsilon_c-i\omega)^{-1/2}}{-i\omega}\,\mathbb E\Big(\frac{\varepsilon_c}{\varepsilon_c-i\omega}\Big),
\label{eq:J}
\end{equation}
where $\mathbb E(m)=\int_0^{\pi/2}\mathrm{d}\theta\,(1-m\sin^2\theta)^{1/2}$. Writing $\mathrm{Im} J(\omega)=\varepsilon_c^{-3/2}j(\omega/\varepsilon_c)$, the crossover function in Eq.~(\ref{eq:dDirH}) is $\mathcal{H}(x)=\int_x^\infty\mathrm{d} u\,j(u)$. The limiting forms $j(u)\simeq2/(\pi u)$ for $u\ll1$ and $j(u)\simeq1/(\sqrt2\,u^{3/2})$ for $u\gg1$ are the two- and one-dimensional return probabilities, and they yield the asymptotes of $\mathcal{H}(x)$ quoted in Sec.~\ref{sec:energy}. The constant of the logarithmic asymptote follows from the expansion of $\mathbb E(m)$ near $m=1$ and was confirmed numerically: $\lim_{x\to0}[\mathcal{H}(x)-(2/\pi)\ln(1/x)]=(8/\pi)\ln2$.

%%%%%%%%%%%%%%%%%%%%%%%%%%%%%%%%%%%%%%%%%%
\begin{adjustwidth}{-\extralength}{0cm}

\reftitle{References}

% Please provide either the correct journal abbreviation (e.g. according to the “The list of Title Word Abbreviations” https://portal.issn.org/ltwa) or the full name of the journal. 
% Citations and references in the Supplementary Materials are permitted provided that they also appear in the reference list here. 

\bibliography{bib}

%=====================================
% References, variant B: internal bibliography
%=====================================

% If authors have biography, please use the format below
%\section*{Short Biography of Authors}
%\bio
%{\raisebox{-0.35cm}{\includegraphics[width=3.5cm,height=5.3cm,clip,keepaspectratio]{Definitions/author1.pdf}}}
%{\textbf{Firstname Lastname} Biography of first author}
%
%\bio
%{\raisebox{-0.35cm}{\includegraphics[width=3.5cm,height=5.3cm,clip,keepaspectratio]{Definitions/author2.jpg}}}
%{\textbf{Firstname Lastname} Biography of second author}

% For the MDPI journals use author-date citation, please follow the formatting guidelines on http://www.mdpi.com/authors/references
% To cite two works by the same author: \citeauthor{ref-journal-1a} (\citeyear{ref-journal-1a}, \citeyear{ref-journal-1b}). This produces: Whittaker (1967, 1975)
% To cite two works by the same author with specific pages: \citeauthor{ref-journal-3a} (\citeyear{ref-journal-3a}, p. 328; \citeyear{ref-journal-3b}, p.475). This produces: Wong (1999, p. 328; 2000, p. 475)

%%%%%%%%%%%%%%%%%%%%%%%%%%%%%%%%%%%%%%%%%%
%% for journal Sci
%\reviewreports{\\
%Reviewer 1 comments and authors’ response\\
%Reviewer 2 comments and authors’ response\\
%Reviewer 3 comments and authors’ response
%}
%%%%%%%%%%%%%%%%%%%%%%%%%%%%%%%%%%%%%%%%%%
\PublishersNote{}
%\isPreprints{}{% This command is only used for ``preprints''.
\end{adjustwidth}
\end{document}